\documentclass[
twocolumn,
prd,
showpacs,
groupedaddress,
superscriptaddress
]{revtex4}
\usepackage{graphicx}
\usepackage{amsmath}
\usepackage{times,mathptmx}

\def\be{\begin{equation}}
\def\ee{\end{equation}}
\def\ba{\begin{eqnarray}}
\def\ea{\end{eqnarray}}
\newcommand{\bp}{\mathbf{p}}
\newcommand{\bk}{\mathbf{k}}
\newcommand{\bq}{\mathbf{q}}
\newcommand{\bv}{\mathbf{v}}
\newcommand{\eV}{\mbox{ \upshape\textrm{eV}}}
\newcommand{\keV}{\mbox{ \upshape\textrm{keV}}}

\newcommand{\MeV}{\mbox{ \upshape\textrm{MeV}}}
\newcommand{\GeV}{\mbox{ \upshape\textrm{GeV}}}

\newcommand{\K}{\mbox{ \upshape\textrm{K}}}

\begin{document}

\title{Searching for dark matter sterile neutrino in laboratory}

\author{Fedor Bezrukov} 
\affiliation{
  Institut de Th\'eorie des Ph\'enom\`enes Physiques,
  Ecole Polytechnique F\'ed\'erale de Lausanne,
  CH-1015 Lausanne, Switzerland}
\affiliation{
  Institute for Nuclear Research of Russian Academy of Sciences,
  Prospect 60-letiya Oktyabrya 7a,
  Moscow 117312, Russia}

\author{Mikhail Shaposhnikov}
\affiliation{
  Institut de Th\'eorie des Ph\'enom\`enes Physiques,
  Ecole Polytechnique F\'ed\'erale de Lausanne,
  CH-1015 Lausanne, Switzerland}

\date{February 9, 2007}

\begin{abstract}
  If the dark matter of the Universe is made of sterile neutrinos
  with the mass in keV region they can be searched for with the help
  of X-ray satellites. We discuss the prospects of \emph{laboratory}
  experiments that can be competitive and complimentary to Space
  missions. We argue that the detailed study of $\beta$ decays of
  tritium and other nuclei with the help of Cold Target Recoil Ion
  Momentum Spectroscopy (COLTRIMS) can potentially enter into
  interesting parameter range and even supersede the current
  astronomical bounds on the properties of dark matter sterile
  neutrino.
\end{abstract}


\maketitle

\section{Introduction.}
The nature of Dark Matter (DM) in the Universe is a puzzle.  Many
different hypothetical particles coming from physics beyond the
Standard Model (SM) were proposed to play a role of dark matter
particle; none of them have been discovered yet.  In this paper we
will discuss possibilities for a laboratory search of one of the dark
matter candidates: sterile neutrino with the mass in keV region.

In short, here is the case for sterile neutrino as a dark matter
particle.  There are not that many experimental facts in particle
physics which cannot be described by the Standard Model.  These are
neutrino oscillations (neutrinos of the SM are exactly massless and do
not oscillate), dark matter (the SM does not have any stable neutral
massive particle) and baryon asymmetry of the Universe (substantial
deviations from thermal equilibrium, needed for baryogenesis, are
absent for experimentally allowed mass of the Higgs boson; in
addition, it is a challenge to use CP violation in
Cabibbo--Kobayashi--Maskawa mixing of quarks to produce baryon
asymmetry in the SM). This calls for an extension of the SM.  Perhaps,
the most economic one, that can describe all these phenomena in a
unified way is the $\nu$MSM of \cite{Asaka:2005an,Asaka:2005pn}.  In
this model three leptonic singlets (other names for them are
right-handed, Majorana or sterile neutrinos) are added, making the
structure of quark and lepton sectors of the theory similar, up to
introduction of Majorana mass for the new leptonic states.  The
Majorana nature of singlet fermions leads to non-zero masses for
active neutrinos and, therefore, to neutrino oscillations, solving in
this way one of the SM problems.  The lightest of these new particles
with the mass in keV region can have a lifetime greater than that of
the Universe \cite{Dolgov:2000ew} and thus can play a role of (warm)
dark matter \cite{Dodelson:1993je}. The preference for keV mass scale
is coming from the cosmological structure formation arguments related
to the missing satellites problem \cite{Moore:1999nt,Bode:2000gq} and
to cuspy DM distributions in cold dark matter cosmologies
\cite{Goerdt:2006rw,Gilmore:2006iy}.  For other astrophysical
applications of keV sterile neutrinos see
\cite{Kusenko:1997sp,Biermann:2006bu,Mapelli:2006ej,Stasielak:2006br,Hidaka:2006sg}.
The presence of two other heavier fermions with the mass in
$\mathcal{O}(1)$~GeV region leads to generation of baryon asymmetry of
the Universe \cite{Asaka:2005pn} via resonant sterile neutrino
oscillations \cite{Akhmedov:1998qx} and electroweak sphalerons
\cite{Kuzmin:1985mm}. These fermions can be searched for in particle
physics experiments with high intensity proton beams
\cite{Shaposhnikov:2006nn,Gorbunov}.

The only way considered up to now to detect the dark matter sterile
neutrino $N$ is through astrophysical X-ray observations
\cite{Dolgov:2000ew,Abazajian:2001vt}. The interaction of $N$ with
intermediate vector bosons $W$ and $Z$ and ordinary charged leptons
and neutrinos is suppressed by the so-called mixing angle
$\theta=m_D/M_M$, where $m_D$ and $M_M$ are respectively the Dirac
and Majorana masses of sterile neutrino.  The main (but undetectable)
decay modes of sterile neutrino are $N\to2\nu+\bar{\nu}$,
$N\to\nu+2\bar{\nu}$.  In addition, sterile neutrino has a radiative
decay channel $N\to\nu+\gamma$, $N\to\bar{\nu}+\gamma$, producing a
narrow line in X-ray spectrum coming from dark matter halos of
different astronomical objects.  Recently a number of constraints on
the mixing angle $\theta$ became available, coming from the analysis
of X-ray data of Chandra and XMM-Newton satellites
\cite{Boyarsky:2005us,Boyarsky:2006zi,Boyarsky:2006fg,Riemer-Sorensen:2006fh,
Watson:2006qb,Riemer-Sorensen:2006pi,Boyarsky:2006ag,Boyarsky:2006kc,
Abazajian:2006jc}.  It is expected that the best results will come
from the analysis of dwarf satellite galaxies in the Milky way halo,
as having the largest mass-to-light ratios and weakest X-ray
background \cite{Boyarsky:2006fg}. 

In addition to X-ray constraints there are bounds on the mass and
momentum of dark matter sterile neutrinos coming from analysis of
Lyman-$\alpha$ forest clouds and structure formation
\cite{Hansen:2001zv,Seljak:2006qw,Viel:2006kd}.  They depend,
however, on specific mechanism of cosmological production of sterile
neutrinos \cite{Asaka:2006ek}. The most conservative limit on the
mass, $M_N>0.3\keV$ is coming from the analysis of rotational curves
of dwarf spheroidal galaxies
\cite{Tremaine:1979we,Lin:1983vq,Dalcanton:2000hn} (Tremaine--Gunn
bound).

Yet another constraint comes from the requirement that the amount of
sterile neutrinos produced in the early universe due to the mixing
with ordinary neutrinos must be smaller than the amount of the dark
matter observed. In the absence of entropy production due to decays
of heavier singlet fermions \cite{Asaka:2006ek} and assuming that the
standard Big Bang theory is valid at temperatures below few hundreds
MeV these bounds are stronger than those coming from X-ray
observations for $M_N < 3.5$~keV \cite{Asaka:2006nq,Asaka:2006rw}.
However, relaxing the above-mentioned assumptions can eliminate these
constraints \cite{Asaka:2006ek,Gelmini:2004ah}.

To summarize, the most conservative constraints on the dark matter
sterile neutrinos are coming from X-ray observations and from
rotational curves of dwarf galaxies; only those will be used in what
follows. In Fig.~\ref{fig:1} we present the main X-ray bounds taken
from \cite{Boyarsky:2006fg,Boyarsky:2006ag} in the mass range
allowed  by the Tremaine--Gunn bound, for comparison with the
proposed laboratory experiment sensitivity. We stress that these
constrains are purely observational and do not depend on any
theoretical bias; the only assumption behind is that the sterile
neutrinos constitute 100\% of the dark matter in the universe. If
only a fraction $p$ of dark matter is in sterile neutrinos, the X-ray
constraints are weaker by a factor of $p$. For $p$ considerably
smaller than $1$ the Tremaine--Gunn bound is also not applicable.

Imagine now that some day an unidentified narrow line will be found
in X-ray observations.  Though there are a number of tests that could
help to distinguish the line coming from DM decays from the lines
associated with atomic transitions in interstellar medium, how can we
be sure that the dark matter particle is indeed discovered?  Clearly,
a laboratory experiment, if possible at all, would play a key role.
Current bounds in the interesting mass region were mostly based on
kink search in beta decay, inspired by possible discovery of 17~keV
neutrino.  The present bounds \cite{Yao:2006px} are given in
Fig.~\ref{fig:1} and are much weaker than required to compete with
X-ray observations.  Fig.~\ref{fig:1} demonstrates that the search
for DM sterile neutrino in terrestrial experiments is very
challenging, as the strength of interaction of DM sterile neutrino
with the matter is roughly $\theta^2$ times weaker than that of
ordinary neutrino. In this paper we analyze different processes where
DM sterile neutrino can be searched for in the laboratory.  We argue
that the only potential possibility is provided by a precise study of
the kinematics of beta decays. We note, in particular, that this kind
of study is not impossible in a view of a novel momentum space
imaging technique (Cold Target Recoil Ion Momentum Spectroscopy,
COLTRIMS), proposed and developed at the end of nineties (for reviews
see \cite{Ullrich1997,Dorner2000}).

The paper is organized as follows.  In second section we will discuss
different reactions in which the DM sterile neutrinos can potentially
manifest themselves.  In  third section we consider the requirements
to the $\beta$-decay recoil experiments that can enter into
interesting region of mixing angle for DM sterile neutrino and
provide arguments that they could be potentially feasible.  The last
section is conclusions.


\section{DM sterile neutrinos in the laboratory.}

In the $\nu$MSM interaction of DM sterile neutrino $N$ with fermions of
the SM can be derived from the standard 4-fermion weak interaction by
replacements
\(
  \nu_\alpha \rightarrow \nu_\alpha +\theta_\alpha N
\),
where $\alpha = e,~\mu,~\tau$, $\theta^2=\sum\theta_\alpha^2$, and we
work in the lowest order in mixing angles $\theta_\alpha$.  Thus,
sterile neutrinos participate in all reactions the ordinary neutrinos
do with a probability suppressed by $\theta^2$.  Additionally, the
fact that they are Majorana particles, $N=N^c$ ($c$ is the sign of
charge conjugation), leads to lepton number non-conservation.

From very general grounds the possible experiments for the search of
sterile neutrinos can be divided in three groups:


{(i)} Sterile neutrinos are \emph{created} and subsequently
\emph{detected} in the laboratory.  The number of events that can be
associated with sterile neutrinos in this case is suppressed by
$\theta^4$ in comparison with similar processes with ordinary
neutrinos.  The smallness of the mixing angle, as required by X-ray
observations, makes this type of experiments hopeless.  For example,
for sterile neutrino mass $m_s=5\keV$, the suppression in comparison
with neutrino reactions is at least of the order of $10^{-19}$.


(ii) Sterile neutrinos are created somewhere else in large amounts
and then \emph{detected} in the laboratory.  The X-ray Space
experiments are exactly of this type: the number density of sterile
neutrinos is fixed by the DM mass density, and the limits on the
X-ray flux give directly the limit on $\theta^2$ rather than
$\theta^4$ as in the previous case.  Another potential possibility is
to look for sterile neutrinos coming from the Sun.  The flux of
sterile neutrinos from, say, $pp$ reactions is
$F_N\sim6\times10^{10}\theta_e^2/\mathrm{cm}^2\mathrm{s}$.  The only
way to distinguish sterile neutrinos from this source from electronic
neutrinos is the kinematics of the reactions $\nu_e n \rightarrow p
e$ and $N n \rightarrow p e$, which looks hopeless.  For higher
energy sources, such as $^8B$ neutrinos, the emission of sterile $N$
would imitate the antineutrinos from the Sun due to the reaction $N p
\rightarrow n e^+$ which is allowed since $N$ is a Majorana particle.
However, this process is contaminated by irremovable background from
atmospheric antineutrinos.  Even if all other sources of background
can be eliminated, an experiment like KamLAND would be able to place
a limit of the order of $\theta^4 < 3\times 10^{-7}$, which is weaker
than the X-ray limit for all possible sterile neutrino masses obeying
the Tremaine--Gunn bound
\cite{Tremaine:1979we,Lin:1983vq,Dalcanton:2000hn}.  The current
KamLAND limit can be extracted from \cite{Eguchi:2003gg} and reads
$\theta^4 < 2.8\times 10^{-4}$.  The sterile neutrino can also be
emitted in supernovae (SNe) explosions in amounts that could be
potentially much larger than $\theta^2 F_\nu$, where $F_\nu$ is the
total number of active neutrinos coming from SNe.  The reason is that
the sterile neutrinos interact much weaker than ordinary $\nu$ and
thus can be emitted from the volume of the star rather than from the
neutrino-sphere.  Using the results of \cite{Dolgov:2000ew}, the flux
of SNe sterile neutrinos due to $\nu_e-N$ mixing is
$
  F_N \simeq 5\times10^{3}\theta_e^2\left(m_s/{\rm keV}\right)^4 F_\nu
$.  
In spite of this enhancement, we do not see any experimental way to
distinguish the $N$ and $(\nu,~\bar\nu)$ induced events in the
laboratory.


(iii) The process of sterile neutrinos \emph{creation} is studied in
the laboratory.  In this case one can distinguish between two
possibilities.  In the first one, we have a reaction which would be
exactly forbidden if sterile neutrinos are absent.  We were able to
find just one process of this type, namely
$S\rightarrow\text{invisible}$, where $S$ is any scalar boson.
Indeed, in the SM the process $S\rightarrow\nu\bar\nu$ is not allowed
due to chirality conservation, and $S\rightarrow\nu\nu$ is forbidden
by the lepton number conservation.  With sterile neutrinos, the
process $S\rightarrow \nu N$ may take place.  However, a simple
estimate shows that the branching ratios for these modes for
available scalar bosons such as $\pi^0$ or $K^0$ are incredibly small
for admitted (by X-ray constraints) mixing angles.  So, only one
option is left out: the detailed study of kinematics of different
$\beta$ decays.

An obvious possibility would be the main pion decay mode
$\pi^-\to\mu\nu$ with creation of sterile neutrino $N$ instead of the
active one.  This is a two body decay, so the energy muon spectrum is
a line with the kinetic energy $(m_\pi-m_\mu)^2/2m_\pi=4.1\MeV$ for
decay with active neutrino and $((m_\pi-m_\mu)^2-m_s^2)/2m_\pi$ for
decay with massive sterile neutrino.  Thus, for $m_s$ of keV order
one needs the pion beam with energy spread less than 0.01~eV to
distinguish the line for sterile neutrino, which seems to be
impossible to get with current experimental techniques.

In the case of beta decay there are two distinct possibilities.  One
is to analyze the electron spectrum only.  In this case the admixture
of sterile neutrinos leads to the kink in the spectrum at the
distance $m_s$ from the endpoint.  However, the distinguishing a
small kink of the order of $\theta^2$ on top of the electron spectra
is very challenging  from the point of view of statistically large
physical background and nontrivial uncertainties in electron spectrum
calculations.

The case of full kinematic reconstruction of beta decay of radioactive
nucleus is more promising and will be analyzed in the next section.


\section{COLTRIMS and $\beta$-decays.}

The idea of using beta decay for sterile neutrino detection is quite
simple: measuring the full kinematic information for the initial
isotope, recoil ion, and electron one can calculate the neutrino
invariant mass on event by event basis.  In an ideal setup of exact
measurement of all these three momenta such an experiment provides a
background-free measurement where a single registered anomalous event
will lead to the positive discovery of heavy sterile neutrino.  This
idea was already exploited at time of neutrino discovery and testing
of the Fermi theory of $\beta$ decay \cite{Pontecorvo1947}.  It was
also proposed to use full kinematic reconstruction to verify the
evidence for 17~keV neutrino found in the kink searches
\cite{Cook:1991cm,Finocchiaro:1992hy} to get rid of possible
systematics deforming the beta spectrum.  Recently, bounds on sterile
neutrino mixing were achieved by full kinematic reconstruction of
$^{38m}$K isotope confined in a magneto--optic trap
\cite{PhysRevLett.90.012501} but for a neutrino in the mass range
$0.7-3.5$~MeV, what is much heavier than considered here. For
$370-640$~keV mass range a similar measurement was performed in
electron capture decay of $^{37}$Ar \cite{Hindi:1998ym}.  We will
discuss below a possible setup for a dedicated experiment for a
search of keV scale DM sterile neutrino.

Let us consider an idealized experiment in which a cloud
of $\beta$-unstable nuclei, cooled to temperature $T$, is
observed.  For example, for $^3$H the normal beta decay is
\[
  {}^3\mathrm{H} \to {}^3\mathrm{He}+e+\bar{\nu}_e\;,
\]
while in presence of sterile neutrino in about $\theta^2$ part of the
events (up to the kinematic factor) the decay proceeds as
\[
  {}^3\mathrm{H} \to {}^3\mathrm{He}+e+N\;,
\]
where $N$ is a sterile neutrino in mostly right-handed helicity
state. Suppose that it is possible to register the recoil momentum of
the daughter ion and of the electron with high enough accuracy. 
Indeed, existing COLTRIMS experiments are able to measure very small
ion recoil \cite{Ullrich1997,Dorner2000}.  They are utilized for
investigation of the dynamics of ionization transitions in atoms and
molecules.  The ion momenta is determined by time of flight
measurement.  A small electric field is applied to the decay region
to extract charged ions into the drift region.  After the drift
region the ions are detected by a position sensitive detector, which
allows to determine both the direction of the momenta and the time of
flight.  Characteristic energies of recoil ion in beta decay is of
the order of the recoil momenta measured by existing COLTRIMS in
ion--atom collisions. Precisions currently achieved with such
apparatus are of the order of $0.2\keV$ for the ion momentum
\cite{Dorner2000,Dorner:97,Mergel:96,Mergel:97,Dorner:98}.

Electron detection is more difficult, as far as the interesting
energy range is of the order of 10~keV for $^3$H decay (or greater
for most other isotopes).  This is much higher than typical energies
obtained in atomic studies.  One possible solution would be to use
the similar time-of-flight technique as for the recoil ions, but with
adding magnetic field parallel to the extraction electric field, thus
allowing to collect electrons from a wider polar angle.  In existing
applications such a method was used for electrons with energies of
only 0.1~keV \cite{Moshammer1996,Kollmus1997}.  In \cite{Jahnke2004}
retarding field was added in the electron drift region allowing to
work with electrons of up to 0.5~keV energies.  Alternatively, one
may try to use electrostatic spectrometers for electron energy
measurement, as it was proposed in
\cite{Cook:1991cm,Finocchiaro:1992hy}.  On the one hand, the latter
method allows to use the electron itself to detect the decay moment
for recoil time of flight measurement. On the other hand, it is hard
to reach high polar angle acceptance with this method, thus losing
statistics.

The decay moment needed for the time-of-flight measurement can be
tagged by registering the Lyman photon emission of the excited ion or
by the electron detection, if electron energy is determined by a
dedicated spectrometer.  Note, that for $^3\mathrm{H}_2$ case Lyman
photon is emitted only in about 25\% of the events
\cite{Finocchiaro:1992hy}, so the photon trigger also induces some
statistics loss.

According to \cite{Dorner2000} it is possible to achieve sensitivity
for measuring normal active neutrino masses of $10\eV$ for each
single event; the accuracy needed for the case of sterile neutrinos
is considerably less than that as the mass of $N$ is expected to be
in the keV region.  Moreover, the measurement in the latter case is a
relative measurement, which is much simpler than absolute measurement
of the peak position required for active neutrino mass determination.

\begin{figure}[t]
\centerline{\includegraphics[width=\columnwidth]{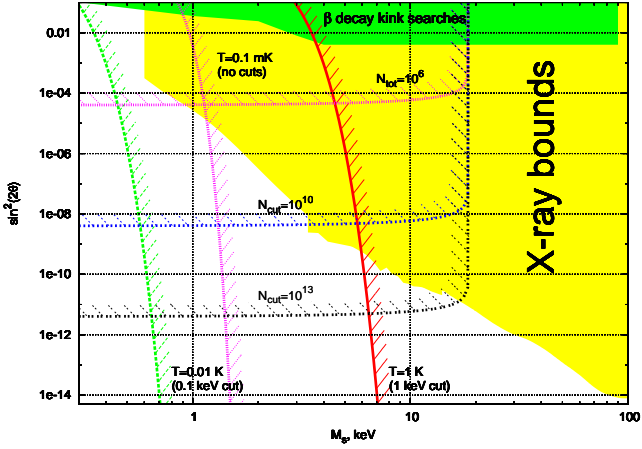}}
\caption{Constraints on the mixing angle $\theta$ of sterile neutrino
  with active neutrino from X-ray observations of Large Magellanic
  Clouds and Milky Way by XMM-Newton and Milky Way by HEAO-1
  satellites and from kink searches in beta decay.  The X-ray bound is
  in the assumption that sterile neutrinos constitute $p=100\%$ of the
  DM, for smaller $p$ the bound is relaxed accordingly.
  The boundaries of the parameter space accessible to $\beta$--decay
  experiments are shown for various values of the source temperature,
  kinematic cuts and collected statistics.}
\label{fig:1}
\end{figure}

Let us estimate the required precision of momentum measurements and
source temperature.  Suppose that an initial molecule with mass $M$
decays \emph{at rest} into recoil ion, electron and neutrino with
momenta $\bp$, $\bk$ and $\bq=\bp+\bk$ respectively. The energy
release will be denoted by $Q$.  Then the neutrino mass can be defined
from ion and electron momenta as
\begin{equation*} 
  m_\nu^2 = (Q-E_e-E_p)^2-(\bp+\bk)^2
  \;,
\end{equation*} 
where $E_e=\sqrt{m_e^2+\bk^2}-m_e$ and $E_p=\sqrt{M^2+\bp^2}-M$ are
electron and recoil ion kinetic energies. It is immediately seen, that
for measuring keV neutrino mass precision of 0.5~keV in momenta
measurement would be sufficient.  For example, for $^3$H decay this
means $0.5\times10^{-2}$ precision in momentum measurement.

Let us turn now to the question of the temperature of the cloud.
Nonzero thermal velocity $\bv$ of the initial molecule spoils the
measurement.  The measured values of the momenta would be
$\tilde{\bp}=\bp+M\bv$ and $\tilde{\bk}=\bk+m_e\bv$ so the measured
mass $m_\nu^{\text{eff}}$ is now
\begin{multline*}
  {m_\nu^{\text{eff}}}^2
    \equiv
    (Q-\tilde{E}_e)^2-(\tilde\bp+\tilde\bk)^2
    \simeq
    m_\nu^2 + M^2\bv^2 - 2M\tilde\bq\bv\;, 
\end{multline*}
where $\tilde\bq\equiv\tilde\bp+\tilde\bk$ and we have neglected terms
of the order of $m_e$, $Q$, and $|\bk|$ compared to the ion mass $M$.
The average squared thermal velocity is $M^2\langle \bv^2\rangle=3MT$,
while the average momenta (in the rest frame) is $|\bq|\lesssim Q$.
It is immediately seen that for reasonably low temperatures the last
term leads to the dominant error. Assuming an isotropic thermal
probability distribution of decaying atoms $P(\bv)\propto
\exp(-M\bv^2/2T)$ it is easy to find (with exponential accuracy) the
probability to get nonzero value of $m_\nu^{\text{eff}}\equiv m_s$
from an event originating from a $\beta$ decay to massless ordinary
neutrino,
\begin{equation}\label{prob}
  P(m_s) \propto \exp\left(
                     -\frac{m_s^2}{2 M T}f^2\left(\frac{|\bq|}{m_s}\right)
                   \right)
  \;,
\end{equation}
where $f(z) = 1/(\sqrt{1+z^2}+z)$.

Comparing this background with the number of sterile neutrino events,
which is proportional to $\theta^2$, we get that the required
temperature must satisfy
\begin{equation*}
  \frac{m_s^2}{2 M T}f^2\left(\frac{|\bq|}{m_s}\right) \gtrsim \log(1/\theta^2)
  \;.
\end{equation*}
Now, if all $\beta$ decay events are considered, the maximal value of
$|\bq|$ is $Q$, and the bound on the temperature is rather rigid and
reads approximatively,
\begin{equation}\label{nocutbound}
  T \lesssim 
    \frac{0.7\times10^{-3}}{\log(1/\theta^2)}\left(\frac{m_s}{1\keV}\right)^4
    \left(\frac{6\GeV}{M}\right)\left(\frac{18.6\keV}{Q}\right)^2
   (1\K)
    \;.
\end{equation}
However, one can loosen this bound considerably at the cost of the
effective source intensity by imposing a kinematic cut on the momenta
$ (\tilde\bq)^2 \lesssim 3MT $, which leads to $|\bq|/m_s\simeq 0$ in
(\ref{prob}) and to much higher acceptable temperatures
\begin{equation}\label{cutbound}
  T \lesssim 
 \frac{1}{\log(1/\theta^2)}\left(\frac{m_s}{1\keV}\right)^2
    \left(\frac{6\GeV}{M}\right)
  (1\K)
    \;.
\end{equation}

Another bound on the experimental sensitivity to the mixing angle is
provided by the requirement that at least one decay with sterile
neutrino happens during the observation.  The number of signal events
is estimated using the differential decay width for massive neutrino
$
  d\Gamma \sim \theta^2 q^2E_e d q
$,
where $E_e\simeq\sqrt{2m_e(Q-\sqrt{q^2+m_\nu^2})}$ is the electron
kinetic energy and $q=|\bq|$ is the neutrino momentum.  If the
kinematic cut on the momenta
is much smaller then $Q$ and $m_s$, the
number of sterile neutrino events can be estimated as
\[
  N_\text{events}
    \simeq
    \theta^2\sqrt{1-({m_s}/{Q})} N_\text{cut}
  \;,
\]
with $N_\text{cut}$ being the number of active neutrino events
satisfying $|\bq|<C$, which is related to the total number of decays
$N_\text{tot}$ by
\[
  \frac{N_\text{cut}}{N_\text{tot}} \simeq
  \frac{35}{16}\left(\frac{C}{Q}\right)^3
  \;.
\]
For the case with generic cut numerical integration of the
differential width can be performed.  One should also note that the
cut on the momentum suppresses active neutrino events more than
sterile neutrino events.

After this general discussion let us turn to specific numbers. As we
see, the best beta decay sources should have small mass and small
energy release.  Then the contribution from the thermal motion of the
decaying atoms will be small and created sterile neutrino will be not
too relativistic, thus making it easier to measure its mass.  Here
the exceptional opportunity is provided with tritium $^3$H, which is
much lighter than all other radioactive elements and has short enough
lifetime of $12.3$ years.  However, if the neutrino mass is higher
than the energy release $Q=18.591 \keV$ in tritium decay other
isotopes should be used. The requirements on the temperature for
the case of $^3\mathrm{H}_2$ molecule can be read directly from
(\ref{nocutbound},\ref{cutbound}) by substituting $M$ by $6\GeV$. 
The bounds (\ref{cutbound}) and $N_\text{events}>10$ ($3\sigma$ with
zero background) for tritium are given on the Fig.~\ref{fig:1} for
several different temperatures and exposures.

One can see, that for $T \sim 0.01 \K$ and $N_\text{cut}\sim 10^{13}$
the experiment is better than astrophysical X-ray bounds for all
achievable for tritium experiment masses.  For a more modest number,
like  $N_\text{cut}\sim 10^{10}$ (i.e.\ year of observation with 1000
decay counts per second, corresponding to a typical recoil ion time
of flight in a 1~m long spectrometer) the sensitivity is smaller, but
still competes with X-ray experiments for a vast range of masses.
Currently \cite{Ullrich2003}, one can create supersonic gas jets with
particle densities of about $10^{11}-10^{12}\mathrm{~cm}^{-3}$ and
temperatures of the order of 0.1~K and prepare sources with
magneto--optical traps with densities of $10^{10}\mathrm{~cm}^{-3}$
and temperature $\sim$ 0.1~mK.  These methods provide $10^6-10^8$
beta decays per year for the source size of about 1~mm$^3$. The line
corresponding to the statistics  $N_\text{tot}\sim 10^{6}$ is shown
in Fig. \ref{fig:1} together with the line for 0.1~mK without making
any cuts; it indicates that the current magneto--optical traps technologies for
low temperature sources allow to obtain bounds better than existing
kink searches and to enter in the interesting parameter region
provided that neutrinos make less than 100\% of DM.

An important point should be emphasized. All the estimates above have
been done with the thermal distribution of the source, which is not
necessarily the case. As far as the number of signal events for keV
neutrino is expected to be extremely small, the tails of the
distribution of thermal velocities are important.  Large non-Gaussian
tails will not spoil too much the precision of the mass measurement,
but will penalize significantly the sensitivity to the mixing angle.
This may be a problem for some cooling techniques.  For example, the
supersonic jet cooling has some non-Gaussian tails in the velocity
distribution \cite{Dorner2000,Miller_ssjets}.


\section{Conclusions.}
In this paper we argued that the detailed study of kinematics of
$\beta$ decays with the help of COLTRIMS may enter into interesting
parameter region for the search of Dark Matter sterile neutrinos
which is complementary to cosmic X-ray missions and indispensable for
revealing the nature of DM in the Universe.  Even with currently
existing technologies entering in the interesting parameter region is
possible for light sterile neutrinos.  Extending the study
to compete with X-ray bounds for higher masses is a challenging but
valuable experimental task.

For a detailed feasibility study of $\beta$-decay experiments to
search for DM sterile neutrino a number of extra points, including
existence of possible backgrounds, should be clarified. One obvious
background appears from the fact that after the $\beta$-decay a
fraction (15\%) of the $(^3\mathrm{He}^3\mathrm{H})^+$ ions
dissociates, leading to an error in the determination of momentum of
the detected recoil ions.  However, the momentum after dissociation is
high \cite{Finocchiaro:1992hy}, so that these events do not look like
the heavy neutrino ones. In addition, one should take into account the
scattering of ions as a source for possible background.  Also, a
careful analysis may lead to the choice of another optimal isotope,
which has higher decay energy release but is short lived, providing
thus larger statistics. A very hard problem is the low density of cold
atoms (serving as a source of beta-decays), available at present.

A similar experiment can be also made with isotopes decaying by
electron capture. In case one has a single-electron ion of such an
isotope, the final state would contain just two particles, and to
find the invisible particle mass ($\nu$ or $N$) it is  sufficient to
measure only the energy of recoil ion, which is a simpler task than
the full momentum reconstruction.  However, if the initial isotope is
not highly ionized, then, generally, one or several Auger electrons
are emitted after the decay, carrying away considerable momentum. 
All these Auger electrons should be detected and taken into account
(see \cite{Hindi:1998ym} for $^{37}$Ar case).  We leave the
comparative study of this type of experiments for future discussion.

\begin{acknowledgments}
  This work was supported in part by the Swiss National Science
  Foundation.  We are indebted to V. A.~Kuzmin for numerous discussions,
  important suggestions and encouragement.  We thank R.~D\"orner for
  valuable comments and D.~Gorbunov and S.~Sibiryakov for helpful
  remarks. Remarks of the referee are greatly appreciated.
\end{acknowledgments}

\bibliographystyle{h-physrev4}
\bibliography{all}

\begin{thebibliography}{10}

\bibitem{Asaka:2005an}
T.~Asaka, S.~Blanchet and M.~Shaposhnikov,
\newblock Phys. Lett. {\bf B631}, 151 (2005), [hep-ph/0503065].

\bibitem{Asaka:2005pn}
T.~Asaka and M.~Shaposhnikov,
\newblock Phys. Lett. {\bf B620}, 17 (2005), [hep-ph/0505013].

\bibitem{Dolgov:2000ew}
A.~D. Dolgov and S.~H. Hansen,
\newblock Astropart. Phys. {\bf 16}, 339 (2002), [hep-ph/0009083].

\bibitem{Dodelson:1993je}
S.~Dodelson and L.~M. Widrow,
\newblock Phys. Rev. Lett. {\bf 72}, 17 (1994), [hep-ph/9303287].

\bibitem{Moore:1999nt}
B.~Moore {\em et~al.},
\newblock Astrophys. J. {\bf 524}, L19 (1999).

\bibitem{Bode:2000gq}
P.~Bode, J.~P. Ostriker and N.~Turok,
\newblock Astrophys. J. {\bf 556}, 93 (2001), [astro-ph/0010389].

\bibitem{Goerdt:2006rw}
T.~Goerdt, B.~Moore, J.~I. Read, J.~Stadel and M.~Zemp,
\newblock Mon. Not. Roy. Astron. Soc. {\bf 368}, 1073 (2006),
  [astro-ph/0601404].

\bibitem{Gilmore:2006iy}
G.~Gilmore {\em et~al.},
\newblock astro-ph/0608528.

\bibitem{Kusenko:1997sp}
A.~Kusenko and G.~Segre,
\newblock Phys. Lett. {\bf B396}, 197 (1997), [hep-ph/9701311].

\bibitem{Biermann:2006bu}
P.~L. Biermann and A.~Kusenko,
\newblock Phys. Rev. Lett. {\bf 96}, 091301 (2006), [astro-ph/0601004].

\bibitem{Mapelli:2006ej}
M.~Mapelli, A.~Ferrara and E.~Pierpaoli,
\newblock Mon. Not. Roy. Astron. Soc. {\bf 369}, 1719 (2006),
  [astro-ph/0603237].

\bibitem{Stasielak:2006br}
J.~Stasielak, P.~L. Biermann and A.~Kusenko,
\newblock Astrophys. J. {\bf 654}, 290 (2007), [astro-ph/0606435].

\bibitem{Hidaka:2006sg}
J.~Hidaka and G.~M. Fuller,
\newblock Phys. Rev. {\bf D74}, 125015 (2006), [astro-ph/0609425].

\bibitem{Akhmedov:1998qx}
E.~K. Akhmedov, V.~A. Rubakov and A.~Y. Smirnov,
\newblock Phys. Rev. Lett. {\bf 81}, 1359 (1998), [hep-ph/9803255].

\bibitem{Kuzmin:1985mm}
V.~A. Kuzmin, V.~A. Rubakov and M.~E. Shaposhnikov,
\newblock Phys. Lett. {\bf B155}, 36 (1985).

\bibitem{Shaposhnikov:2006nn}
M.~Shaposhnikov,
\newblock Nucl. Phys. {\bf B763}, 49 (2007), [hep-ph/0605047].

\bibitem{Gorbunov}
D.~Gorbunov and M.~Shaposhnikov,
\newblock in preparation.

\bibitem{Abazajian:2001vt}
K.~Abazajian, G.~M. Fuller and W.~H. Tucker,
\newblock Astrophys. J. {\bf 562}, 593 (2001), [astro-ph/0106002].

\bibitem{Boyarsky:2005us}
A.~Boyarsky, A.~Neronov, O.~Ruchayskiy and M.~Shaposhnikov,
\newblock Mon. Not. Roy. Astron. Soc. {\bf 370}, 213 (2006),
  [astro-ph/0512509].

\bibitem{Boyarsky:2006zi}
A.~Boyarsky, A.~Neronov, O.~Ruchayskiy and M.~Shaposhnikov,
\newblock Phys. Rev. {\bf D74}, 103506 (2006), [astro-ph/0603368].

\bibitem{Boyarsky:2006fg}
A.~Boyarsky, A.~Neronov, O.~Ruchayskiy, M.~Shaposhnikov and I.~Tkachev,
\newblock Phys. Rev. Lett. {\bf 97}, 261302 (2006), [astro-ph/0603660].

\bibitem{Riemer-Sorensen:2006fh}
S.~Riemer-Sorensen, S.~H. Hansen and K.~Pedersen,
\newblock Astrophys. J. {\bf 644}, L33 (2006), [astro-ph/0603661].

\bibitem{Watson:2006qb}
C.~R. Watson, J.~F. Beacom, H.~Yuksel and T.~P. Walker,
\newblock Phys. Rev. {\bf D74}, 033009 (2006), [astro-ph/0605424].

\bibitem{Riemer-Sorensen:2006pi}
S.~Riemer-Sorensen, K.~Pedersen, S.~H. Hansen and H.~Dahle,
\newblock astro-ph/0610034.

\bibitem{Boyarsky:2006ag}
A.~Boyarsky, J.~Nevalainen and O.~Ruchayskiy,
\newblock astro-ph/0610961.

\bibitem{Boyarsky:2006kc}
A.~Boyarsky, O.~Ruchayskiy and M.~Markevitch,
\newblock astro-ph/0611168.

\bibitem{Abazajian:2006jc}
K.~N. Abazajian, M.~Markevitch, S.~M. Koushiappas and R.~C. Hickox,
\newblock astro-ph/0611144.

\bibitem{Hansen:2001zv}
S.~H. Hansen, J.~Lesgourgues, S.~Pastor and J.~Silk,
\newblock Mon. Not. Roy. Astron. Soc. {\bf 333}, 544 (2002),
  [astro-ph/0106108].

\bibitem{Seljak:2006qw}
U.~Seljak, A.~Makarov, P.~McDonald and H.~Trac,
\newblock Phys. Rev. Lett. {\bf 97}, 191303 (2006), [astro-ph/0602430].

\bibitem{Viel:2006kd}
M.~Viel, J.~Lesgourgues, M.~G. Haehnelt, S.~Matarrese and A.~Riotto,
\newblock Phys. Rev. Lett. {\bf 97}, 071301 (2006), [astro-ph/0605706].

\bibitem{Asaka:2006ek}
T.~Asaka, M.~Shaposhnikov and A.~Kusenko,
\newblock Phys. Lett. {\bf B638}, 401 (2006), [hep-ph/0602150].

\bibitem{Tremaine:1979we}
S.~Tremaine and J.~E. Gunn,
\newblock Phys. Rev. Lett. {\bf 42}, 407 (1979).

\bibitem{Lin:1983vq}
D.~N.~C. Lin and S.~M. Faber,
\newblock Astrophys. J. {\bf 266}, L21 (1983).

\bibitem{Dalcanton:2000hn}
J.~J. Dalcanton and C.~J. Hogan,
\newblock Astrophys. J. {\bf 561}, 35 (2001), [astro-ph/0004381].

\bibitem{Asaka:2006nq}
T.~Asaka, M.~Laine and M.~Shaposhnikov,
\newblock JHEP {\bf 0701}, 091 (2007), [hep-ph/0612182].

\bibitem{Asaka:2006rw}
T.~Asaka, M.~Laine and M.~Shaposhnikov,
\newblock JHEP {\bf 06}, 053 (2006), [hep-ph/0605209].

\bibitem{Gelmini:2004ah}
G.~Gelmini, S.~Palomares-Ruiz and S.~Pascoli,
\newblock Phys. Rev. Lett. {\bf 93}, 081302 (2004), [astro-ph/0403323].

\bibitem{Yao:2006px}
Particle Data Group, W.~M. Yao {\em et~al.},
\newblock J. Phys. {\bf G33}, 1 (2006).

\bibitem{Dorner2000}
R.~Dorner {\em et~al.},
\newblock Physics Reports {\bf 330}, 95 (2000).

\bibitem{Ullrich1997}
J.~Ullrich {\em et~al.},
\newblock Journal of Physics B: Atomic, Molecular and Optical Physics {\bf 30},
  2917 (1997).

\bibitem{Eguchi:2003gg}
KamLAND, K.~Eguchi {\em et~al.},
\newblock Phys. Rev. Lett. {\bf 92}, 071301 (2004), [hep-ex/0310047].

\bibitem{Pontecorvo1947}
B.~Pontecorvo,
\newblock Reports on Progress in Physics {\bf 11}, 32 (1947).

\bibitem{Cook:1991cm}
S.~Cook, M.~Fink, S.~D. Thomas and H.~Wellenstein,
\newblock Phys. Rev. {\bf D46}, R6 (1992).

\bibitem{Finocchiaro:1992hy}
G.~Finocchiaro and R.~E. Shrock,
\newblock Phys. Rev. {\bf D46}, R888 (1992).

\bibitem{PhysRevLett.90.012501}
M.~Trinczek {\em et~al.},
\newblock Phys. Rev. Lett. {\bf 90}, 012501 (2003).

\bibitem{Hindi:1998ym}
M.~M. Hindi {\em et~al.},
\newblock Phys. Rev. {\bf C58}, 2512 (1998).

\bibitem{Dorner:97}
R.~D\"orner {\em et~al.},
\newblock Nucl. Instr. and Meth. B {\bf 124}, 225 (1997).

\bibitem{Mergel:96}
V.~Mergel,
\newblock PhD thesis, Universit\"at Frankfurt, Shaker, Aachen, 1996.

\bibitem{Mergel:97}
V.~Mergel {\em et~al.},
\newblock Phys. Rev. Lett. {\bf 79}, 387 (1997).

\bibitem{Dorner:98}
R.~D\"orner {\em et~al.},
\newblock Phys. Rev. A {\bf 57}, 1074 (1998).

\bibitem{Moshammer1996}
R.~Moshammer, J.~Ullrich, M.~Unverzagt, W.~Schmitt and B.~Schmidt-Bocking,
\newblock Nuclear Instruments and Methods in Physics Research Section B: Beam
  Interactions with Materials and Atoms {\bf 108}, 425 (1996).

\bibitem{Kollmus1997}
H.~Kollmus, W.~Schmitt, R.~Moshammer, M.~Unverzagt and J.~Ulrich,
\newblock Nuclear Instruments and Methods in Physics Research Section B: Beam
  Interactions with Materials and Atoms {\bf 124}, 377 (1997).

\bibitem{Jahnke2004}
T.~Jahnke {\em et~al.},
\newblock Journal of Electron Spectroscopy and Related Phenomena {\bf 141}, 229
  (2004).

\bibitem{Ullrich2003}
J.~Ullrich {\em et~al.},
\newblock Reports on Progress in Physics {\bf 66}, 1463 (2003).

\bibitem{Miller_ssjets}
D.~Miller,
\newblock in {\em Atomic and Molecular Beam Methods} Vol.~14, Oxford University
  Press, New York, 1988.

\end{thebibliography}

\end{document}